\documentclass[amsmath, amssymb,10pt, aps, prb, reprint, notitlepage, longbibliography, superscriptaddress]{revtex4-2}

\usepackage{graphicx}
\usepackage{dcolumn}
\usepackage{hyperref}
\usepackage{bm}
\usepackage{xcolor}
\usepackage{braket}
\usepackage{tabularx}
\usepackage{booktabs}
\usepackage{makecell}
\usepackage{multirow}
\usepackage{relsize}
\usepackage{bibunits}
\usepackage{comment} 
\defaultbibliography{references}
\defaultbibliographystyle{apsrev4-2}

\begin{document}
\begin{bibunit}[apsrev4-2] 

\title{Low-temperature entropies and possible states in geometrically frustrated magnets}   

\date{\today}

\author{Siyu Zhu}
\affiliation{Physics Department, University of California, Santa Cruz, California 95064, USA}

\author{Arthur P. Ramirez}
\affiliation{Physics Department, University of California, Santa Cruz, California 95064, USA}

\author{Sergey Syzranov}
\affiliation{Physics Department, University of California, Santa Cruz, California 95064, USA}

\begin{abstract}
    The entropy that an insulating magnetic material releases upon cooling
    can reveal important information about the 
    properties of spin states in that material.
    In many geometrically frustrated (GF) magnetic compounds, 
    the heat capacity exhibits a low-temperature peak that comes from the spin states
    continuously connected to the ground states of classical models,
    such as the Ising model, on the same GF lattice, which manifests 
    in the amount of entropy associated with this heat-capacity peak.
    In this work, we simulate numerically the values of entropy released
    by higher-spin triangular-lattice layered systems and materials on
    SCGO lattices.    
    We also compare the experimentally measured values of entropy
    in several strongly GF compounds, $NiGa_2S_4$, $FeAl_2Se_4$ and SCGO/BSZCGO,
    with possible theoretical values inferred from 
    the classical models to which the quantum states of those materials may be connected.
    This comparison suggests that the lowest-energy states 
    of higher-spin layered triangular-lattice compounds can be described in terms of doublet states on individual magnetic sites.
    Our analyses demonstrate how the values of entropy can reveal the structure
    of low-energy magnetic states in GF compounds and call for more accurate thermodynamic measurement in GF magnetic materials.
\end{abstract}

\maketitle

\section{Introduction}

\label{Sec:Introduction}

The amount of entropy released by a magnetic material upon cooling may reveal fundamental properties of spin states in that material. For example,
if a magnetic material is cooled down from high temperatures, at which 
it behaves as free spins, to $T=0$ and releases the amount of entropy smaller than the entropy $S_\text{free}=N\ln(2S+1)$ of free spins,
this suggests the extensive degeneracy of the ground state, specific to spin ices~\cite{Ramirez:SpinIceEntropy} and certain
spin-liquid models~\cite{SavaryBalents:review,KnolleMoessner:review}. 

\begin{figure}[b]
\centering
\includegraphics[width=0.85\linewidth]{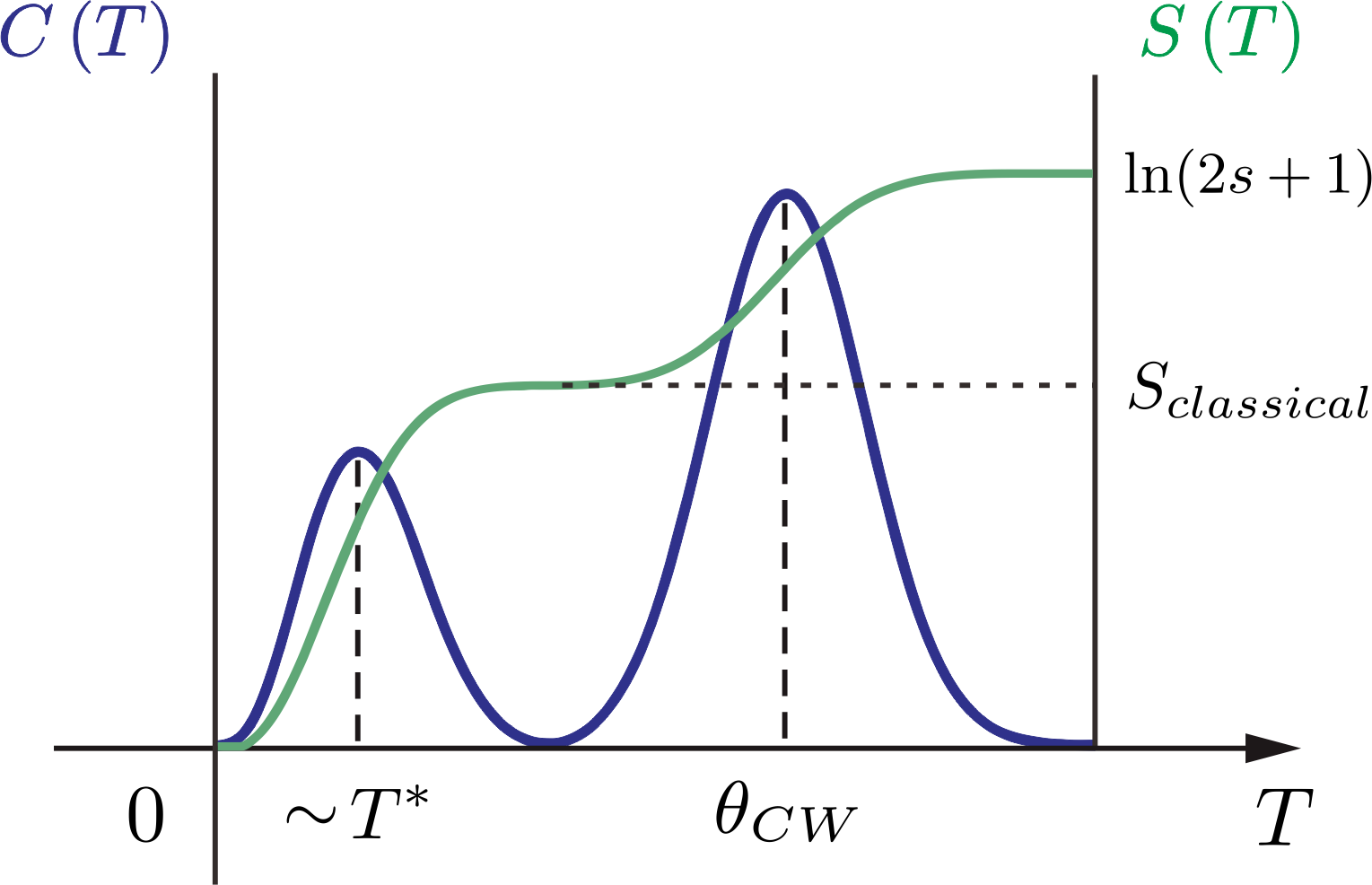}
\caption{\label{fig:C_peak}
The heat capacity and the entropy per spin
(in units of the ideal gas constant $R$) as a function of temperature in
a geometrically frustrated magnet.
The heat capacity exhibits two peaks, near the ``hidden energy scale'' $T^*$,
and near the Curie-Weiss temperature $\theta_{CW}$.
The peaks may partially overlap in some GF compounds.
The total entropy $S=\int_0^\infty C(T)/T dT$ per spin associated with the two peaks
is given by the entropy $\ln(2s+1)$ of a free spin.
The entropy of the lower peak matches the ground-state 
entropy $S_\text{classical}$ of a classical (e.g. Ising) spin model
on the same lattice.
}
\end{figure}

The amount of entropy may also reveal the structure of the 
low-energy states of a magnetic material and its characteristic 
energy scales. GF magnets have recently been predicted~\cite{PoppRamirezSyzranov,RamirezSyzranov:GFreview} to be continuously
connected to classical Ising models. The states connected 
to the Ising ground states lead to the formation of a low-temperature peak in the behavior of the specific heat, as shown in Fig.~\ref{fig:C_peak}.

Spin-$1/2$ XXZ models on the kagome lattice have been known ~\cite{Elser:KHAF,ZengElser:KHAF,IsodaNakano:XXZ} for decades
to exhibit two distinct heat-capacity peaks,
near the ``hidden energy scale'' $T^*$~\cite{Syzranov:HiddenEnergy}
and near the Curie-Weiss temperature $\theta_{CW}$, with the corresponding states
connected, respectively, to the Ising ground states
and to spin-flip excitations. Recently, such a 
structure of the specific heat has been shown to be a generic 
feature of GF magnetic materials~\cite{PoppRamirezSyzranov,RamirezSyzranov:GFreview}.

Hereafter, the states of two models are said to be 
continuously connected if they evolve into each other when one model is continuously
deformed into the other.
For example, the classical Ising model is deformed to the XXZ model
with the Hamiltonian, 
\begin{align}
    H_{XXZ}=-J\sum_{\langle i,j\rangle} S_i^z S_j^z
    -J_\perp
    \sum_{\langle i,j\rangle} \left(S_i^x S_j^x+S_i^y S_j^y\right),
    \label{Hxxz}
\end{align}
which describes a broad class of GF magnets, by
increasing the transverse spin-spin coupling $J_\perp$ from zero to a finite value.

The realistic Hamiltonian of a material may be significantly more complicated than
the Hamiltonian~\eqref{Hxxz} of the XXZ model and include a number of components with
unknown parameters: dipole-dipole interactions, quenched disorder, single-ion anisotropy, etc. The entropy $S_\text{peak}=\int_\text{peak}\frac{C(T)}{T}dT$ associated with the specific-heat peaks
will, however,
be robust against such ingredients so long as the heat-capacity
peaks remain well separated
from each other.

The entropy values of classical Ising ground states - 
the values that may manifest themselves
in quantum materials with various unknown, complicated microscopic details - 
are known for most common lattices,
as summarized in Table~\ref{tab:spin_lattice}. 
\begin{table}[h!]
\centering
\begin{tabular}{|c|c|c|}
\hline
\textbf{Spin} & \textbf{Lattice} & \textbf{Ground-state entropy} \\
\hline
\multirow{4}{*}{
spin $\frac{1}{2}$} & Triangular & 0.323066~\cite{Wannier:Ising,Wannier:erratum, SedikSiyuSergey:Vacancy} \\
\cline{2-3}
& Kagome     & 0.50183~\cite{Kenzi:KagomeEntropy} \\
\cline{2-3}
& Pyrochlore   & 0.205507~\cite{Singh:pyrochloreEntropy, Pauling:SpinIce, Nagle:SpinIce, Ramirez:SpinIceEntropy} \\
\cline{2-3}
& Hyperkagome & 0.502~\cite{Pohle:HyperkagomeEntropy, Wills:HyperkagomeEntropy} \\
\hline
spin 1 & Triangular & 0.435854~\cite{Zukovic:spin1triangular}, [This work]  \\
\hline
\multirow{2}{*}{
spin $\frac{3}{2}$} & SCGO  &  \multirow{2}{*}{ 0.331991 [This work]}  \\
& B(SZ)CGO &   \\
\hline
\end{tabular}
\caption{
Ground-state entropies (per spin) (in units of the ideal gas constant $R$)
for the antiferromagnetic
Ising models and their higher-spin counterparts for several common frustrating lattices.
}
\label{tab:spin_lattice}
\end{table}
For a number of commonly investigated GF magnets, however, the theoretical values 
still await analytical or numerical calculations.

To investigate possible values of entropies in GF materials,
we carry out numerical simulations of the ground-state entropies of the 
spin-$1$ equivalent of the antiferromagnetic Ising model on the triangular lattice
and the spin-$3/2$ equivalents of the Ising model on the lattices of  
$SrCr_{9}Ga_{3}O_{19}$, $BaCr_{9}Ga_{3}O_{19}$, and
$Ba_2Sn_2ZnCr_{7}Ga_{3}O_{22}$,
known, respectively, as SCGO~\cite{Ramirez:SCGOentropy}, BCGO~\cite{Yang:BCGO},
and BSZCGO~\cite{Piyakulworawat:BSZCGO,HagemannCava:BaSnGaZnCrO}.

{We emphasize that the listed materials are not described by the
Ising models, nor have Ising models with spins $s>\frac{1}{2}$ have ever been observed~\cite{RamirezSyzranov:GFreview}. Nevertheless, comparing 
the values of entropy measured in materials with the 
(theoretical) values of entropy for Ising models
can give valuable insights about the nature of the low-energy states of those materials, as they may be continuously connected to the ground states of 
Ising models.
}

In some compounds, with several examples studied in this paper, 
the quantum states of GF magnets may also be continuously connected to 
other classical models, for example, 
Potts or Blume-Capel models.
Identifying the low-energy ground states of GF magnetic materials is further complicated by the fact that
the effective low-energy magnetic degrees of freedom may correspond to lower values of 
spins than the atomic spins of the compound due
to magnetocrystalline anisotropy
and spin-orbit hybridization. 

In this paper, we analyze possible values of the magnetic entropy that GF magnetic 
materials may exhibit and compare those values with the
entropy values measured in several materials thoroughly studied in experiments.
Such a comparison allows us to identify the effective degrees of freedom and 
possible structures of low-energy states in several GF compounds.

The paper is organized as follows. In Sec.~\ref{Sec:numerics},
we present our numerical results for the ground-state entropies in the higher-spin
counterparts of the antiferromagnetic Ising models on the triangular and 
SCGO lattices.
In Secs.~\ref{Sec:TriangularCompounds1} and \ref{Sec:SCGOCompounds2},
we compared the expected entropy values in, respectively,
triangular-lattice and SCGO-type compounds with the values of entropy
measured in experiments.
We conclude in Sec.~\ref{Sec:Conclusion}.


\section{Numerical results for Ising ground-state entropies}

\label{Sec:numerics}

In this section, we provide numerical results for the 
ground-state entropies in the Ising models
and higher-spin counterparts of Ising models on several common
frustrating lattices.
As discussed in Sec.~\ref{Sec:Introduction}, the respective ground-state
entropies are expected to match the entropies 
associated with the low-temperature heat-capacity peaks in quantum materials
on the same lattice, which may be 
described by the Heisenberg XXZ or more complicated quantum models.


\medskip
\noindent\textit{Spins-1 on a triangular lattice.} 
Two-dimensional spin-$1/2$ Ising models allow for analytical calculations of their
ground-state entropies~\cite{kac1952combinatorial,Landafshitz5,Wannier:Ising,Wannier:erratum, Kenzi:KagomeEntropy,SedikSiyuSergey:Vacancy}. 
Entropies of higher spins on GF lattices cannot be computed analytically in a similar way and 
require a numerical evaluation.

The ground-state entropies of higher-spin counterparts of the Ising model, hereafter referred 
to as higher-spin Ising models, have been computed numerically in Ref.~\cite{Zukovic:spin1triangular}. Integrating the quantity $C(T)/T$
obtained by means of Monte-Carlo simulations using the Metropolis algorithm,
the ground-state entropy in the spin-$1$ Ising model has been found to be
\begin{align}
    \tilde{S}_1=0.43472\pm 0.00004.
\end{align}

In this paper, we independently verify the value of the ground-state entropy for the 
spin-$1$ Ising model using the Wang-Landau algorithm, augmented by the adaptive $1/t$ modification~\cite{Belardinelli:t_Algor, Belardinelli:t_Algor1} (see 
Appendix~\ref{Spin1triangular} for details). The result of our simulations,
\begin{align}
    S_{1} = 0.435854 \pm 0.000030,
\end{align}
is fairly close to the result obtained in Ref.~\cite{Zukovic:spin1triangular}
using a different method.


\medskip
\noindent\textit{Spins-$3/2$ on the SCGO (BCGO) lattice.} 
In what immeidately follows, we explore the ground-state entropy of the spin-$3/2$
Ising model on the lattice of the 
well-known frustrated magnet SrCr$_{9}$Ga$_{3}$O$_{19}$ 
(SCGO)~\cite{Ramirez:SCGOentropy} and its isomorphic relative
BaCr$_{9}$Ga$_{3}$O$_{19}$ (BCGO)~\cite{Yang:BCGO}.

\begin{figure}[hb!]
\centering
\includegraphics[width=1\linewidth]{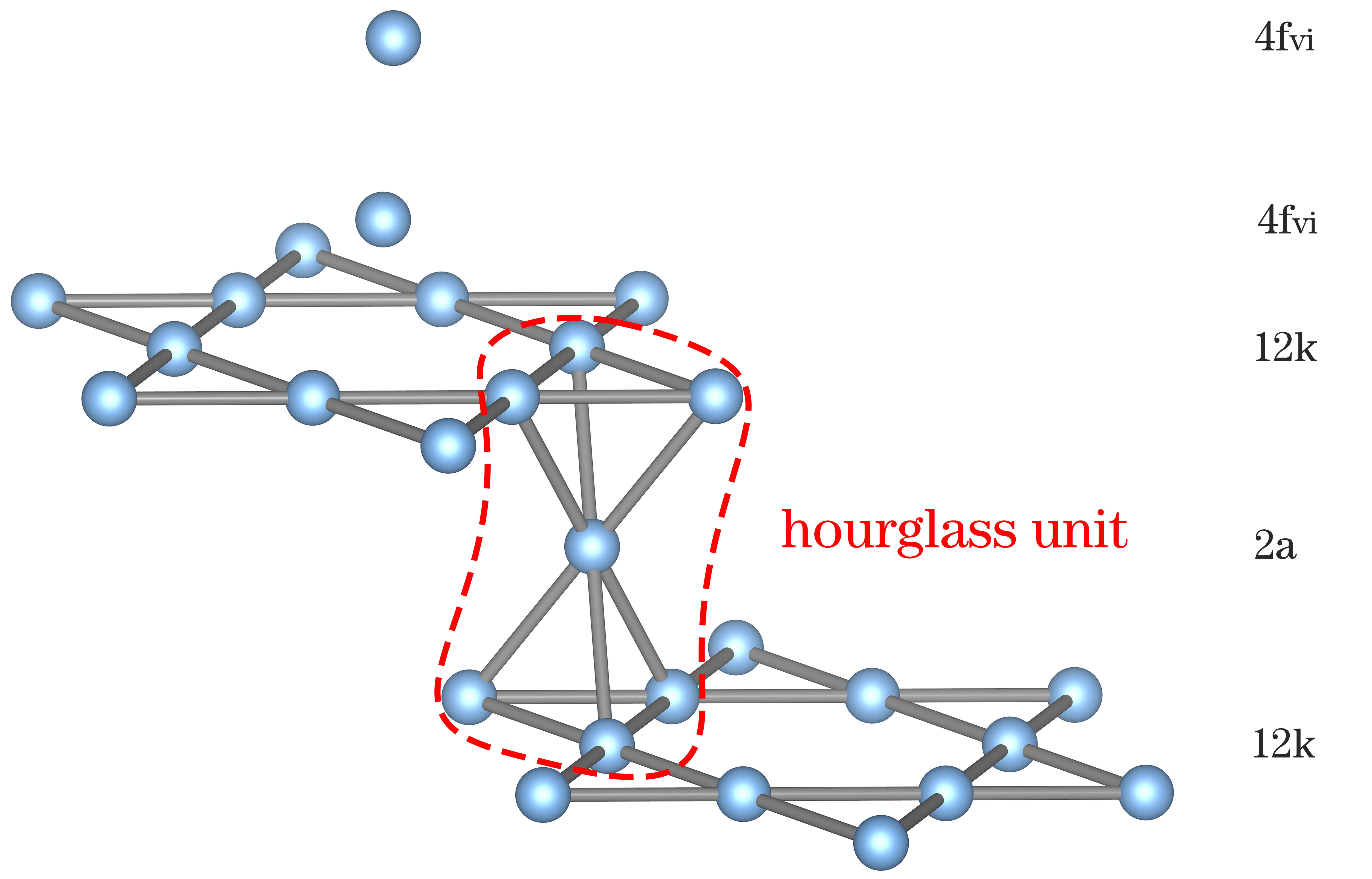}
\caption{\label{fig:SCGO}
Cr$^{3+}$ sites in SCGO. 
The labels on the right (4f$_{\mathrm{vi}}$, 12k, 2a) denote the crystallographic Wyckoff positions 
of the Cr$^{3+}$ ions in the magnetoplumbite structure.
The kagome layers correspond to the 12k sites, the triangular layers to the 2a sites, and the interlayer ions to the 4f$_{\mathrm{vi}}$ sites. 
The red dashed outline highlights an \emph{hourglass unit}, consisting of
two corner-sharing tetrahedra (12k–2a–12k).
The BCGO and BSZCGO compounds have a similar magnetic structure but do not contain 
the 4f$_{\mathrm{vi}}$ sites.
}
\end{figure}

Both materials are characterized by a layered geometry composed of Cr$^{3+}$ ions forming corner-sharing triangles—realizing kagome–triangle–kagome trilayers—with the magnetic ions occupying three distinct crystallographic sites: $12k$ (kagome layer), $2a$ (triangular layer), and $4f_\mathrm{vi}$ (interlayer dimers), as shown in Fig.~\ref{fig:SCGO}.

Although SCGO contains additional spins in the $4f_\mathrm{vi}$ planes (see Fig.~\ref{fig:SCGO}), 
those spins have been shown to be decoupled from the 
rest of the spins at low temperatures~\cite{Lee:SCGO_4f}. 
Those planes have also been demonstrated to host singlet spins sates. 
Depending on the fraction of the $4f_\mathrm{vi}$ spins, which we will
discuss in Sec.~\ref{Sec:SCGOCompounds2},
the contribution of such spins to entropy can be obtained analytically.

In this section, we focus, therefore, 
on the remaining spin network consisting 
of the layers of touching pyramids on the $12k-2a-12k$ sites (which can also be viewed as a layer of hourglass units shown in Fig.~\ref{fig:SCGO}).
 This same magnetic backbone is present in BCGO, with $Ba$ replacing $Sr$. Neutron diffraction
data~\cite{Yang:BCGO} show that BCGO is isostructural to SCGO, with minor modifications such as a slightly expanded $c$-axis and larger $4f_\mathrm{vi}$–$4f_\mathrm{vi}$ bond distances.

To determine the ground-state entropy of the frustrated lattice shared by SCGO and BCGO, we find numerically the ground states of a layer of hourglass units of classical spins-$3/2$, i.e. variables that can take four values, $S_i = \pm 3/2, \pm 1/2$.
Hereinafter, we refer to this model as the spin-$3/2$ Ising model 
on the SCGO lattice.

We apply the Wang–Landau algorithm with the $1/t$ modification~\cite{WangLaudau:WL, WangLaudau:WL1, Belardinelli:t_Algor, Belardinelli:t_Algor1} to compute the density of states and obtain the ground-state entropy. We simulate lattice sizes ranging from $2 \times 2$ to $21 \times 21$ hourglass units (corresponding to $28$ to $3087$ spins) under periodic boundary conditions.

\begin{figure}[hb!]
\centering
\includegraphics[width=1\linewidth]{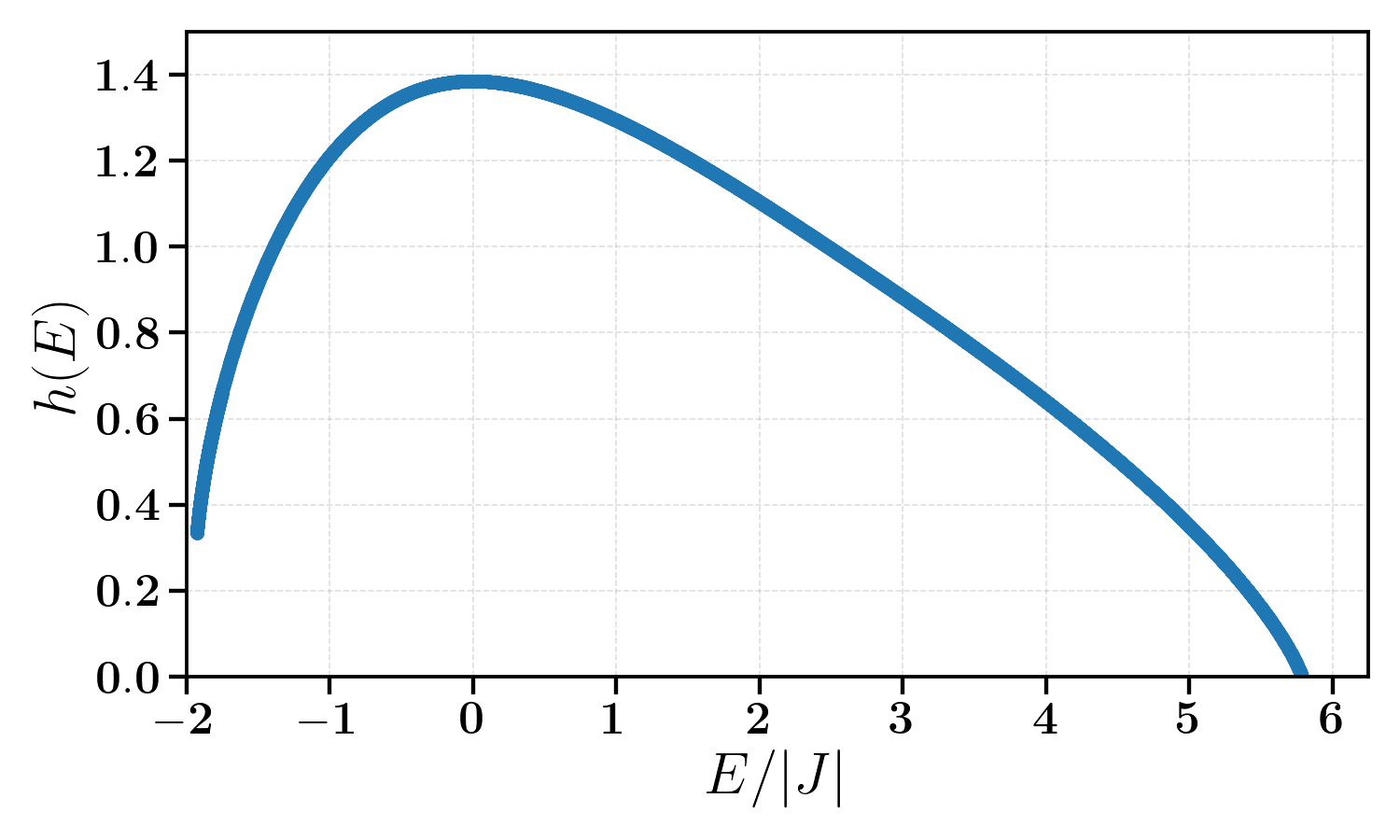}
\caption{\label{fig:SCGO_DOS}
The microcanonical entropy $h(E)=\ln g(E)/N$ per spin, where 
$g(E)$ is the degeneracy of the system's level with the energy $E$ (per spin),
in the spin-3/2 Ising model on the kagome–triangle–kagome trilayer lattice, characteristic of SCGO and BCGO, consisting of $21\times 21$
hourglass units shown in Fig.~\ref{fig:SCGO}.
}
\end{figure}

{
The resulting microcanonical entropy $h(E)=\ln g(E)/N$ per spin, where $g(E)$ is the
degeneracy of the level with energy $E$, is shown in Fig.~\ref{fig:SCGO_DOS}.
At the lowest energy, the quantity $h(E)$ gives the ground-state entropy. 
At the highest energy, all spins are collinear, and $h(E)=0$.
}

\begin{widetext}

\begin{figure}[h!]
\centering
\includegraphics[width=0.9\linewidth]{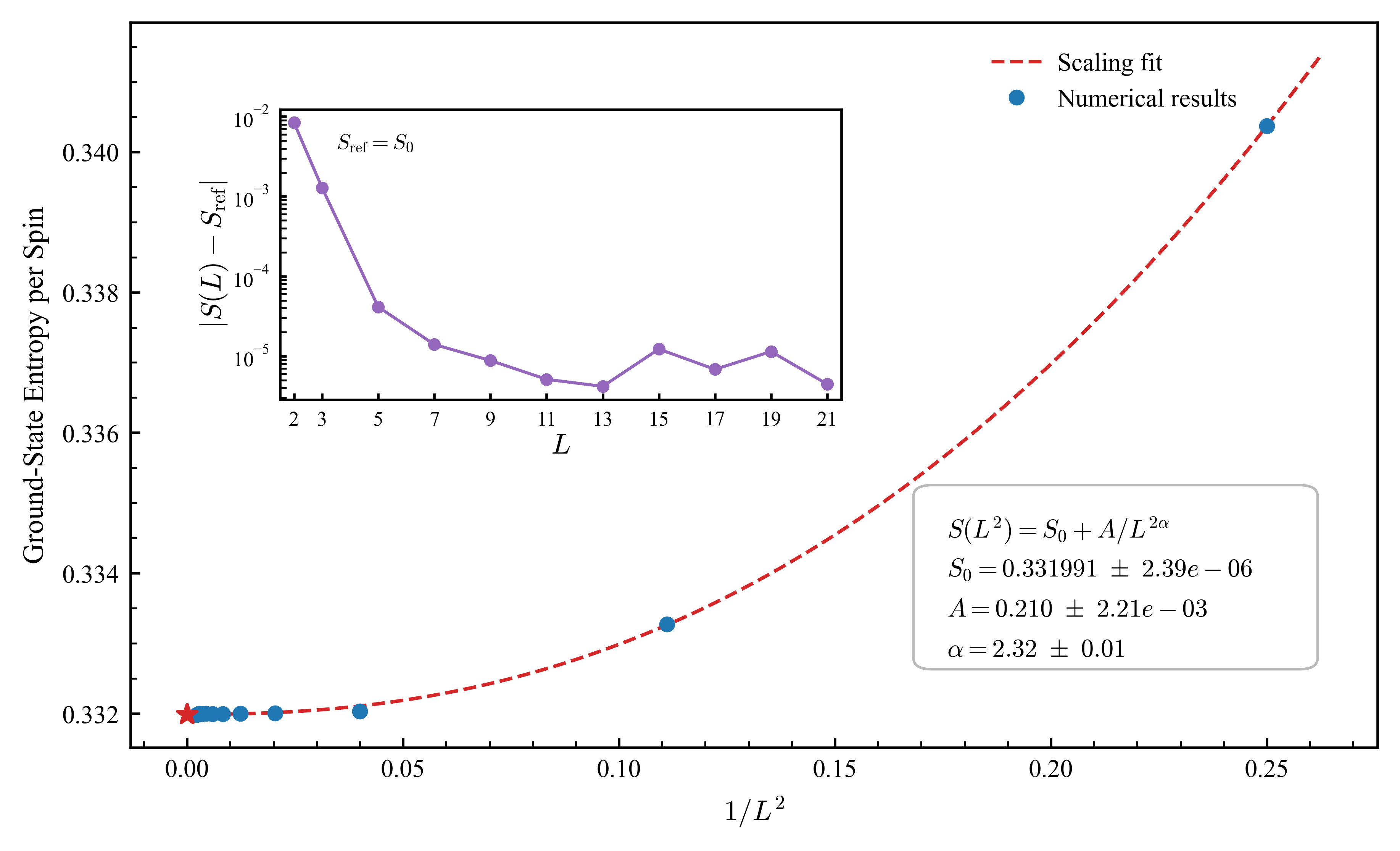}
\caption{\label{fig:SCGOresult}
Non-linear fit results for the ground-state entropy per spin of the SCGO/BCGO lattice.
Where $L$ is the linear dimension of the system, and $N = L^2$ is the total hourglass units.
The entropy per spin converges rapidly and stabilizes around the thermodynamic limit of approximately $0.331991$ per spin. The numerical errors are smaller than the symbol size and therefore not shown.
}
\end{figure}

\end{widetext}

As shown in Fig.~\ref{fig:SCGOresult}, the ground-state entropy per spin exhibits rapid convergence as the system size increases. Only minimal fluctuations are beyond the size of $7\times7$, ensuring a reliable extrapolation to the thermodynamic limit.

Our simulations yield the thermodynamic-limit estimate of the ground-state entropy
\begin{equation}
\label{}
\begin{aligned}
S_{}^{\text{SCGO}} = 0.331991 \pm 0.000002
\end{aligned}
\end{equation}
per spin (in units of the ideal gas constant $R$).

{The uncertainty reflects the standard error of the extrapolated value $S^{\text{SCGO}}$ obtained from a weighted nonlinear least-squares fit of the finite-size data to $S(L) = S_0 + A/L^{2\alpha}$, where $S_0$, $A$ and $\alpha$ are fitting parameters, and each data point is weighted by the standard deviation from multiple independent simulations. The uncertainty is taken from the square root of the corresponding diagonal element of the covariance matrix returned by the fit.
}

\section{Analysis of measured entropies:
triangular-lattice compounds}

\label{Sec:TriangularCompounds1}

In what follows, we compare 
the experimentally measured values of entropy associated with
specific-heat peaks in particular materials
with theoretical values of entropy in various models.
While the microscopic details of the materials are in general poorly known,
as discussed in Sec.~\ref{Sec:Introduction},
such a comparison may shed light on the structure of the states
and the effective degrees of freedom at low energies.

In this section, we focus on the analyses of triangular-lattice layered materials
$NiGa_2S_4$ and $FeAl_2Se_4$. The measured values of the entropy in those materials
and theoretical expectations are summarized in Table~\ref{tab:TrinagularCompounds}.
As detailed below, the theoretically expected values are based on the (classical) Ising models
to which the lowest-energy states of the discussed (quantum) materials may be continuously connected. We show that although magnetism in these materials comes from higher spins
($S=1$ in $NiGa_2S_4$ and $S=2$ in $FeAl_2Se_4$), the data suggest that
their lowest-energy states are continuously connected to the 
ground states of the spin-$1/2$ Ising model on the triangular lattice.

\begin{table}[h!]
    \centering
    \begin{tabular}{|c|c|c|c|}
    \hline
       \textbf{Material} & \makecell{\textbf{Measured}\\ \textbf{entropy}} & \makecell{\textbf{Theoretical} \\ \textbf{expectation}} & \textbf{Classical model} \\
    \hline
     \multirow{2}{*}{$NiGa_2S_4$}    & \multirow{2}{*}{$0.35$~\cite{NakatshujiMaeno:NaGaSneutron}} & $0.32\ldots$ & Spin-$1/2$
      \\
     \cline{3-4}
      & & $0.435\ldots$ & Spin-$1$ \\
     \hline
     \multirow{4}{*}{$FeAl_2Se_4$} &  & $0.32\ldots$ & Spin-$1/2$\\
     \cline{3-4}
        &  \multirow{2}{*}{0.10~\cite{Li:FeAl2Se4}} & $0.435\ldots$ & Spin-$1$\\
     \cline{3-4}
        &  & $0.11\ldots$ & \makecell{Difference between \\
        Spin-$1$ and Spin-$1/2$}\\
    \hline
    \end{tabular}
    \caption{A summary of the values of entropy (per spin) (in units of the ideal gas
    constant $R$)
    associated with the low-temperature heat-capacity peaks
    in higher-spin triangular-lattice
    compounds and the corresponding 
    theoretical expectations based on several models with effective spins.
    The ``Classical model'' refers to an Ising model or its higher-spin counterpart to which 
    the low-energy states of the materials may be continuously connected.}
    \label{tab:TrinagularCompounds}
\end{table}


\subsection{$NiGa_2S_4$}

The layered, triangular-lattice, spin $S=1$ antiferromagnet $NiGa_2S_4$ 
is a good testbed for investigating the connection between the quantum
ground state of a geometrically frustrated magnet and an appropriate classical model.
This material displays two well-separated peaks in the heat capacity $C(T)$ that has been measured to good accuracy in Refs.~\cite{NakatshujiMaeno:NaGaSneutron} and
\cite{Nambu:NiGaS}.

The entropy per spin associated with the lowest-temperature peak 
measured in experiment is approximately $0.35$~\cite{NakatshujiMaeno:NaGaSneutron} (hereafter indicated
in units of the ideal gas constant $R$).
This value is noticeably below the ground-state entropy
$S_1=0.43585\ldots$ of the spin-$1$ Ising model obtained numerically
in this paper. It is close, however, to the
ground-state entropy
$S_{1/2}=0.323066\ldots$ of the spin-$1/2$ Ising model on the triangular lattice~\cite{Wannier:Ising,Wannier:erratum} (cf. Table~\ref{tab:TrinagularCompounds}).

This suggests that, despite being a spin-$1$ magnet,
$NiGa_2S_4$ has a manifold of lowest-energy states
continuously connected to the  ground states of the 
Ising model described by lower spins, i.e. spins-$1/2$.

A plausible scenario for the emergence of such states is the existence of 
perturbations, on top of the spin-spin exchange interactions, that  break the
spin-rotation symmetry and
lift the energy of one of the states of the spin-$1$ on each site away
from the lowest-energy states that give rise to the low-temperature
peak of the specific heat.

The microscopic model describing such a compound can be continuously connected to a subset of states of the  
Blume-Capel model with a negative single-ion anisotropy term ($\Delta<0$), with the Hamiltonian given by
\begin{equation}
\label{Hamton2}
\begin{aligned}
\mathcal{H} = -J\sum_{\langle i,j\rangle} S_i^z S_j^z
+\Delta \left(S_i^z\right)^2,
\end{aligned}
\end{equation}
where $J<0$ and $S_i^z$ is the $z$-component of the a spin-$1$ on site $i$.

Such a model has an extensively degenerate ground state with the entropy of the antiferromagnetic spin-$1/2$ Ising model. When continuously deforming the model, e.g.,
the transverse coupling $\propto -J_\perp
\sum_{\langle i,j\rangle} \left(S_i^x S_j^x+S_i^y S_j^y\right)$ can lift
the degeneracy of the ground states and lead to the formation of the lower-temperature
peak observed in the specific heat of the material.
Continuous deformations of the model do not alter the entropy associated with
the low-temperature peak, so long as the peak remains well separated from the 
higher-temperature features of the specific heat.

To sum up, while the determination of the exact microscopic Hamiltonian 
describing $NiGa_2S_4$ requires further experimental progress,
our analysis suggests that this spin-$1$ magnet
is described by effective spin-$1/2$ degrees of freedom,
with the states continuously
connected to the ground states of the 
antiferromagnetic (spin-$1/2$) Ising model on the triangular lattice.


\subsection{$FeAl_2Se_4$}

$ FeAl_2Se_4$ represents another intriguing 
GF magnet on a triangular lattice~\cite{Li:FeAl2Se4}.
Magnetism in this material comes from the $S=2$ spins of the 
$Fe^{2+}$ ions hybridized with their orbital states~\cite{AbragamBleaney1970}.

The material exhibits two distinct, well-separated peaks, at approximately $10K$ and $65K$,
in the temperature dependencies of the magnetic contribution to the specific heat~\cite{Li:FeAl2Se4}.
By digitizing the data reported in Ref.~\cite{Li:FeAl2Se4}, 
we find the value of entropy of $0.10$
associated with the lower-temperature peak.
This value 
is significantly smaller than the 
ground-state entropy $S_{1/2}=0.323\ldots$ of the spin-$1/2$
Ising model and its higher-spin counterparts (see Table~\ref{tab:TrinagularCompounds}).

The features of the specific heat $C(T)$ in $FeAl_2Se_4$  can be understood as follows.
Spin-orbital hybridization in the material splits the spin states
into a triplet, a quintet, and a septet, with the triplet states having 
the lowest energies~\cite{AbragamBleaney1970}. 

The observed entropy of the lowest-temperature peak in $C(T)$
is close to the entropy difference $S_1-S_{1/2}$ between the ground state entropies
of the spin-$1$ and spin-$1/2$ Ising models ($S_1=0.43585\ldots$ and $S_{1/2}=0.323066\ldots$).
This suggests that the lowest-energy states of the material are continuously connected
to the spin-$1/2$ Ising model, while at higher temperatures ($T\gtrsim 20K$),
the states of the system are connected to those of the spin-$1$ Ising model
(Blume-Capel model with $\Delta=0$).

Similarly to the case of $NiGa_2S_4$, such a structure of the low-energy states 
may appear in the presence of a 
perturbation with 
the characteristic energy $E_\text{anisotr}\sim 10K$ that would further split the triplet
states discussed above, lifting one of the respective states and making the low-energy degrees of freedom of the material equivalent to spins-$1/2$.
At temperatures exceeding $E_\text{anisotr}$, the perturbation is not significant,
and the considered triplet states are equivalent to an effective spin-$1$.
The described scenario can be further verified by low-temperature thermodynamic measurements and neutron-scattering experiments.


\section{SCGO-type compounds}

\label{Sec:SCGOCompounds2}

\begin{table}[h!]
    \centering
    \begin{tabular}{|c|c|c|}
    \hline
       \textbf{Material} & \makecell{\textbf{Measured}\\ \textbf{entropy}} & \textbf{Theoretical prediction} \\
    \hline
     \multirow{2}{*}{SCGO}    & \multirow{2}{*}{$0.565$~\cite{Ramirez:SCGOentropy}} & \makecell{$0.566\ldots$ \\ assuming free $4f_\mathrm{vi}$ spins}
      \\
     \cline{3-3}
      & & \makecell{$0.332\ldots$ \\ if all $4f_\mathrm{vi}$ spins form singlets} \\
     \hline
     BSZCGO    &  $0.62$~\cite{Piyakulworawat:BSZCGO} & $0.332\ldots$\\
    \hline
     BCGO & Not measured & $0.332\ldots$ \\
    \hline
    \end{tabular}
    \caption{A summary of the values of entropy (per spin)
    (in units of the ideal gas constant $R$)
    associated with the low-temperature heat-capacity peaks
    in the SCGO-family compounds and the theoretical predictions we make in this paper.}
    \label{tab:SCGOcompounds}
\end{table}

In the layered GF magnetic materials
$SrCr_{9}Ga_{3}O_{19}$ (SCGO), $BaCr_{9}Ga_{3}O_{19}$ (BCGO), and
$Ba_2Sn_2ZnCr_{7}Ga_{3}O_{22}$ (BSZCGO),
magnetism comes from Cr$^{3+}$ ions with spin $S=3/2$. While the specific-heat 
data necessary for identifying entropy in SCGO and BSZCGO is available in, respectively,
Refs.~\cite{Ramirez:SCGOentropy} and \cite{Piyakulworawat:BSZCGO}, similar 
thermodynamic measurements still remain to be carried out for BCGO.
In what follows, we analyze the experimentally observed entropies associated 
with the lower-temperature heat-capacity peaks in SCGO and BSZCGO.
A summary of the experimentally measured values of entropy together with our
theoretical predictions is given in Table~\ref{tab:SCGOcompounds}.

{\bf SCGO.} The heat capacity of SCGO comes from both 
the hourglass units in each bilayer of pyramids
and, additionally, from the spins of the $4f_\mathrm{vi}$ electrons located between
the bi-pyramid layers shown in Fig.~\ref{fig:SCGO}.
If the Cr ions in the $4f_\mathrm{vi}$ layers were completely free spins-$3/2$ ($S=3/2$), they would contribute an entropy of
$\ln 4 $ per spin. Given that the Cr$^{3+}$ ions in the $4f_\mathrm{vi}$ layers
constitute $2/9$ of all Cr$^{3+}$ ions in the compound, and the rest $7/9$ of the 
Cr$^{3+}$ ions are located in the hourglass units, the expected entropy of the lower-temperature peak per spin in SCGO
is given by
\begin{align}
    \label{SCGOexpectationEntropy}
    \frac{2}{9}\times\ln 4 + \frac{7}{9}\times 0.332 = 0.5663,
\end{align}
where $0.332$ is the computed in Sec.~\ref{Sec:numerics} value of the ground-state Ising entropy per spin in the bi-pyramid structure.

Heat capacity $C(T)$ of the compound $SrCr_{9p}Ga_{12-9p}O_{19}$ with $p=0.98$
in the temperature interval $T=3-100K$ has been reported in Ref.~\cite{Ramirez:SCGOentropy}.
The material has been shown to exhibit a peak in the behaviour of $C(T)$ with the 
entropy $S_{\mathrm{exp}}\approx0.565$ per Cr$^{3+}$ ion, which is very close to the theoretical expectation~\eqref{SCGOexpectationEntropy}.

We emphasize, however, that
while the prediction~\eqref{SCGOexpectationEntropy} is based on the assumption of free $4f_\mathrm{vi}$ spins
in SCGO, not all
of these spins are free.
The neutron scattering experiment~\cite{Lee:SCGO_4f} provides evidence 
of the presence of singlet states of the Cr$^{3+}$
spins in the $4f_{\mathrm{vi}}$ layers.
The measured singlet-triplet gap is 
approximately $18.6$~meV,
corresponding to a temperature of $215K$,
significantly exceeding the  
temperature range
of the heat-capacity peak observed in Ref.~\cite{Ramirez:SCGOentropy}.

A plausible scenario consistent with both the 
thermodynamic~\cite{Ramirez:SCGOentropy} and neutron-scattering~\cite{Lee:SCGO_4f} measurements consists in only a small fraction
of the $4f_\mathrm{vi}$ spins being
in the singlet states, while the rest of them behaving as free spins (while possibly interacting with the spins in the bi-pyramid layer shown in 
Fig.~\ref{fig:SCGO}). In this scenario, the few singlet-state $4f_\mathrm{vi}$ spins
exhibit the singlet-triplet gap observed via neutron 
scattering~\cite{Lee:SCGO_4f}, while the other, effectively free $4f_\mathrm{vi}$
spins together with the bi-pyriamid spins 
lead to an entropy close to the theoretical expectation~\eqref{SCGOexpectationEntropy}.


{\bf BSZCGO.} In contrast to SCGO~\cite{Ramirez:SCGOentropy}, the BSZCGO~\cite{Piyakulworawat:BSZCGO} compound does not contain the additional $4f_{\mathrm{vi}}$ Cr sites; its magnetism originates purely from the bilayer kagome network of Cr$^{3+}$ ions~\cite{Piyakulworawat:BSZCGO} shown in 
Fig.~\ref{fig:SCGO}. The theoretical expectation for the entropy (per Cr$^{3+}$ ion)
under the lower-temperature
peak in the heat capacity of this material is given 
by the numerical result for the ground-state entropy $S_{\infty}^{\text{SCGO}}\approx 0.332$
of the Ising model on the respective lattice obtained in Sec.~\ref{Sec:numerics}.

In Ref.~\cite{Piyakulworawat:BSZCGO}, a peak was observed in the temperature dependence
of the heat capacity in the temperature interval $T=0.5-50K$.
The entropy under the peak is given by $\Delta S\approx 0.45\times
\ln 4\approx 0.62$ per Cr$^{3+}$ spin, which significantly
exceeds the expected value of approximately $0.33$.
The reason for the discrepancy between the theoretical expectation
for the value of entropy and the reported experimental value
remains to be investigated.

The missing entropy of $\ln 4- \Delta S \approx 0.66$ per spin is attributed in Ref.~\cite{Piyakulworawat:BSZCGO} to the zero-point
entropy, i.e. the entropy of the ground state in BSZCGO.
We believe, however, that this missing entropy is most likely
located at the higher-temperature peak that is positioned at temperatures
$T>50 K$ outside of the range of temperatures accessed in Ref.~\cite{Piyakulworawat:BSZCGO}.
The presence of such a second peak in the behavior of the heat capacity is a generic feature of strongly GF magnetic materials~\cite{PoppRamirezSyzranov,RamirezSyzranov:GFreview}.

In agreement with the above picture,
one of us previously showed~\cite{HagemannCava:BaSnGaZnCrO} that the Curie-Weiss constant in clean BSZCGO is $312 K$. We, therefore, expect significant entropy and the associated peak
in the $300K$ range, which has not yet been accessed experimentally.
The presence of such a higher-temperature peak remains to be observed explicitly in both SCGO and BSZCGO.

Special attention must be also given to a careful verification of the entropy associated with the lower-temperature peak in $C(T)/T$.
In Ref.~\cite{Piyakulworawat:BSZCGO}, to obtain the magnetic contribution to the specific heat of BSZCGO,
some care is taken to subtract the specific heat of its non-magnetic isomorph above $50K$
from the total specific heat of the material, but a substantial uncertainty still exists in the region of temperatures where both values of the specific heat are small.

Ironically, even though $C(T)/T$ has a maximum in the region
of low temperatures near the ``hidden energy scale'' $T^*$~\cite{Syzranov:HiddenEnergy,PoppRamirezSyzranov,RamirezSyzranov:GFreview},
accurate determination of the integral
$\int_\text{lower peak}\frac{C(T)}{T}dT$
requires the knowledge of the entropy released at significantly higher temperatures, where a simple subtraction of a non-magnetic isomorph's $C(T)/T$ becomes technically challenging, exemplified
by SCGO and BSZCGO. 
The material's non-ideal attributes, such as long-range interactions or in this case, vestigial magnetic atoms, can introduce experimental artifacts that further complicate the identification of low energy excitations.


\section{Conclusion}

\label{Sec:Conclusion}

We have analyzed the values of entropy in geometrically frustrated (GF) compounds 
on several common lattices. 
Such compounds are known~\cite{RamirezSyzranov:GFreview} to generically exhibit a double-peak structure 
in the temperature dependence $C(T)$ of the specific heat

The entropy associated with the lower-temperature peak 
in a GF compound is expected to match the ground-state entropy of the appropriate Ising
or other classical models on the magnetic lattice of the respective compound.
This value of entropy is robust; it is insensitive to the microscopic details such as 
dipole-dipole interactions, quenched disorder, and single-ion anisotropy, so long as the
$C(T)$ peaks remain well-separated.

{\it Triangular-lattice compounds.}
In higher-spin triangular lattice compounds (spin-$1$ $NiGa_2S_4$ and spin-$2$ $FeAl_2Se_4$),
the measured values of entropy suggest the emergence of effective spin-$1/2$
degrees of freedom, as the entropies associated with lowest-temperature
heat-capacity peaks match either the ground-state entropy of the (spin-$1/2$) Ising model
or the difference between the ground-state entropy of the spin-1/2 Ising model and 
the spin-$1$ model. 

While the analyzed compounds have higher spins, the 
effective spin-$1/2$ degrees of freedom may emerge as a result of 
the splitting of the manifold of the higher-spin states by, e.g., spin-orbit
interactions and single-ion anisotropy.
The details of the microscopic Hamiltonians leading to such effective degrees of freedom
can be further investigated using such probes as
fluorescence spectra and neutron-scattering experiments.

{\it SCGO-type compounds.}
We carried out the numerical simulations of the spin-$3/2$ counterpart of the Ising
on the SCGO lattice, characteristic of the family of compounds including  
$SrCr_{9}Ga_{3}O_{19}$ (SCGO), $BaCr_{9}Ga_{3}O_{19}$ (BCGO), and
$Ba_2Sn_2ZnCr_{7}Ga_{3}O_{22}$ (BSZCGO). As with other GF materials with well-separated 
heat-capacity peaks, the respective value of the ground-state Ising entropy
is expected to match the entropy associated with the lowest-temperature peak in the heat capacity.

Our theoretical expectation for the entropy of the low-temperature peak in SCGO
is in good agreement with the experimental measurements~\cite{Ramirez:SCGOentropy}, assuming that the $4f_\mathrm{vi}$ spins,
which constitute $2/9$ of all spins in this material, act as free spins. 
Although singlet formation among these spins has been observed~\cite{Lee:SCGO_4f}, the measured entropy remains consistent with our predictions if the singlet fraction is small. The microscopic nature of these $4f_\mathrm{vi}$ states therefore warrants further investigation.

Unlike the case of SCGO, the measured entropy~\cite{Piyakulworawat:BSZCGO}
associated with the low-temperature peak in BSZCGO is not in agreement
with the theoretical expectation, exceeding the
latter by approximately a factor of $2$. The reason for this discrepancy calls for further
exploration.

Existing measurements of the heat capacity in SCGO and BSZCGO have so far captured only one peak in the heat capacity.
Based on the generic theory of GF magnetic materials~\cite{PoppRamirezSyzranov,RamirezSyzranov:GFreview},
we expect the existence of the second peak in the heat capacity
of those materials at temperatures of the order of the Curie-Weiss constants,
respectively, $500K$ and $312K$, which so far have been outside of the temperature
ranges accessed in experiments. 

The data described in this paper emphasizes the need for further thermodynamic measurements of the thermodynamic properties of the SCGO, BCGO and BSZCGO compounds in the high-temperature range,
as well as for additional accurate measurements of the low-temperature
heat capacity of GF materials. We have demonstrated how the value of the entropy associated with 
the heat-capacity peaks in such materials can reveal the structure of their low-energy states
and the effective degrees of freedom in them.


\section{Acknowledgements}

This work has been supported by the NSF grant DMR2218130. 
We are also grateful to an anonymous PRL referee for bringing Ref.~\cite{Li:FeAl2Se4}
to our attention.
The numerical simulations reported in this paper were carried out using resources of the shared high performance computing facility at the University of California, Santa Cruz (UCSC). Support for the 
Hummingbird computational cluster
is provided by Information Technology Services Division and the Office of Research at UCSC.



%

\end{bibunit}



\begin{bibunit}[apsrev4-2]
\newpage
\vspace{2cm}
\twocolumngrid

\cleardoublepage

\setcounter{page}{1}

\renewcommand{\theequation}{S\arabic{equation}}
\renewcommand{\thefigure}{S\arabic{figure}}
\renewcommand{\thetable}{S\arabic{table}}
\renewcommand{\bibnumfmt}[1]{[S#1]}
\renewcommand{\citenumfont}[1]{S#1}

\setcounter{equation}{0}
\setcounter{figure}{0}
\setcounter{enumiv}{0}

\appendix


\section{Ground-state entropy of spin-$1$ Ising model on the triangular lattice}

\label{Spin1triangular}

In this section, we provide the details of the numerical simulations of the 
entropy of the spin-$1$ equivalent
of the antiferromagnetic Ising model on the triangular lattice.
Such a model is known as the Blume–Capel model~\cite{Blume:Blume_Capel1, Capel:Blume_Capel2, Capel:Blume_Capel3, Jozef:Blume_Capel4}
with zero crystal‑field term ($\Delta=0$).
The Hamiltonian of the model is given by
\begin{equation}
\label{Hamton}
\begin{aligned}
\mathcal{H} = -J\sum_{\langle i,j\rangle} S_i S_j,
\end{aligned}
\end{equation}
where each variable $S_i$ takes values ${-1,0,+1}$, and $J<0$ corresponds to an antiferromagnetic coupling. 

\begin{widetext}

\begin{figure}[h]
	\centering
	\includegraphics[width=0.9\linewidth]{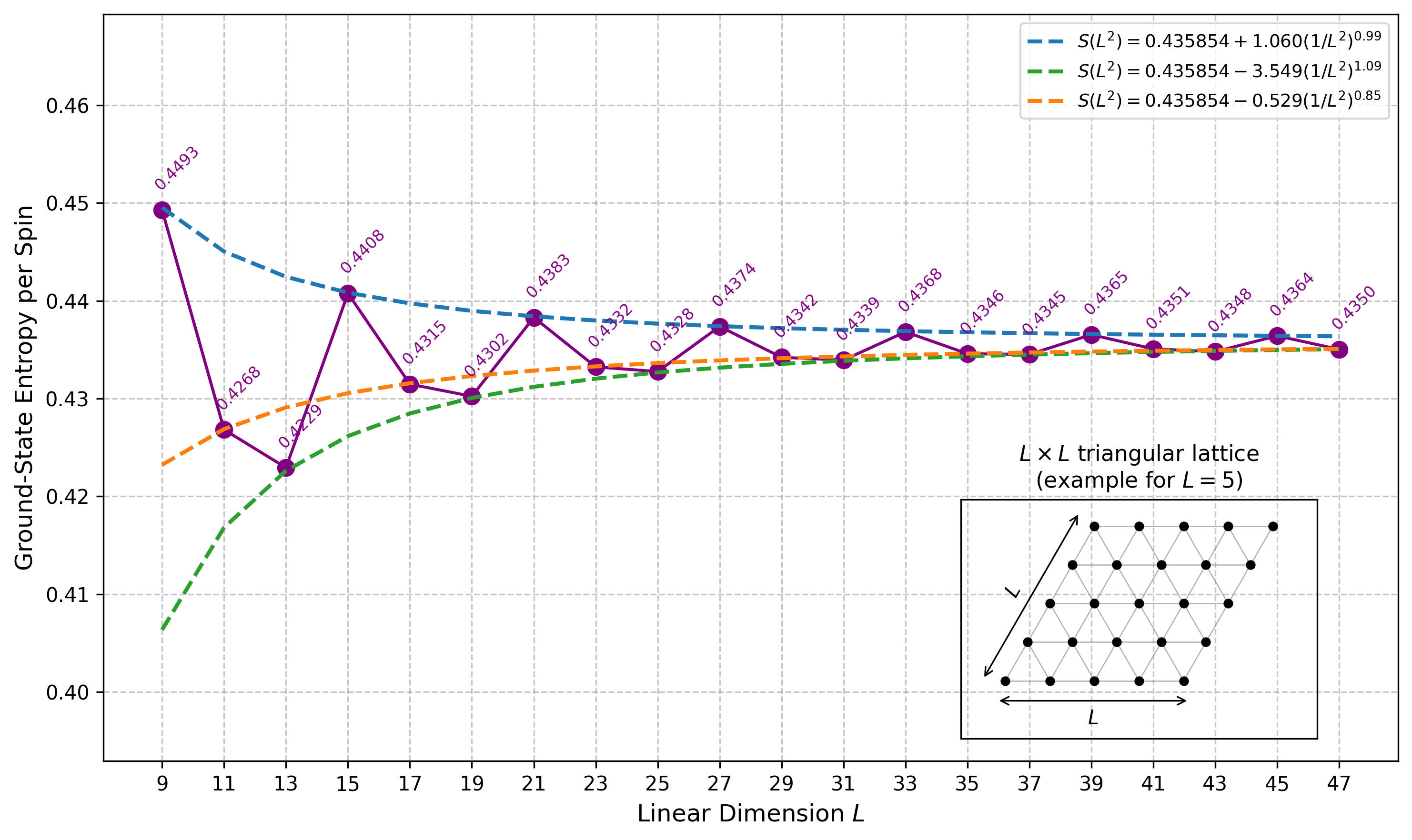}
	\caption{\label{fig:TriEntropy} 
    The ground-state entropy (per spin) in the spin-1 Ising model on a triangular lattice of size $L\times L$ as a function of $L$.
     }
\end{figure}

\end{widetext}

We carry out numerical simulations of the ground-state entropy in the model described by the Hamiltonian~\eqref{Hamton} on the triangular lattice with periodic boundary conditions (PBC) (see the Appendix~\ref{PBC explanation} for details).
We simulate systems that have $L$ cells along each of the two directions
of the Bravais lattice (see Fig.~\ref{fig:TriEntropy}), here referred
to  as $L\times L$-size systems. 

\begin{figure}[h]
	\centering
	\includegraphics[width=1\linewidth]{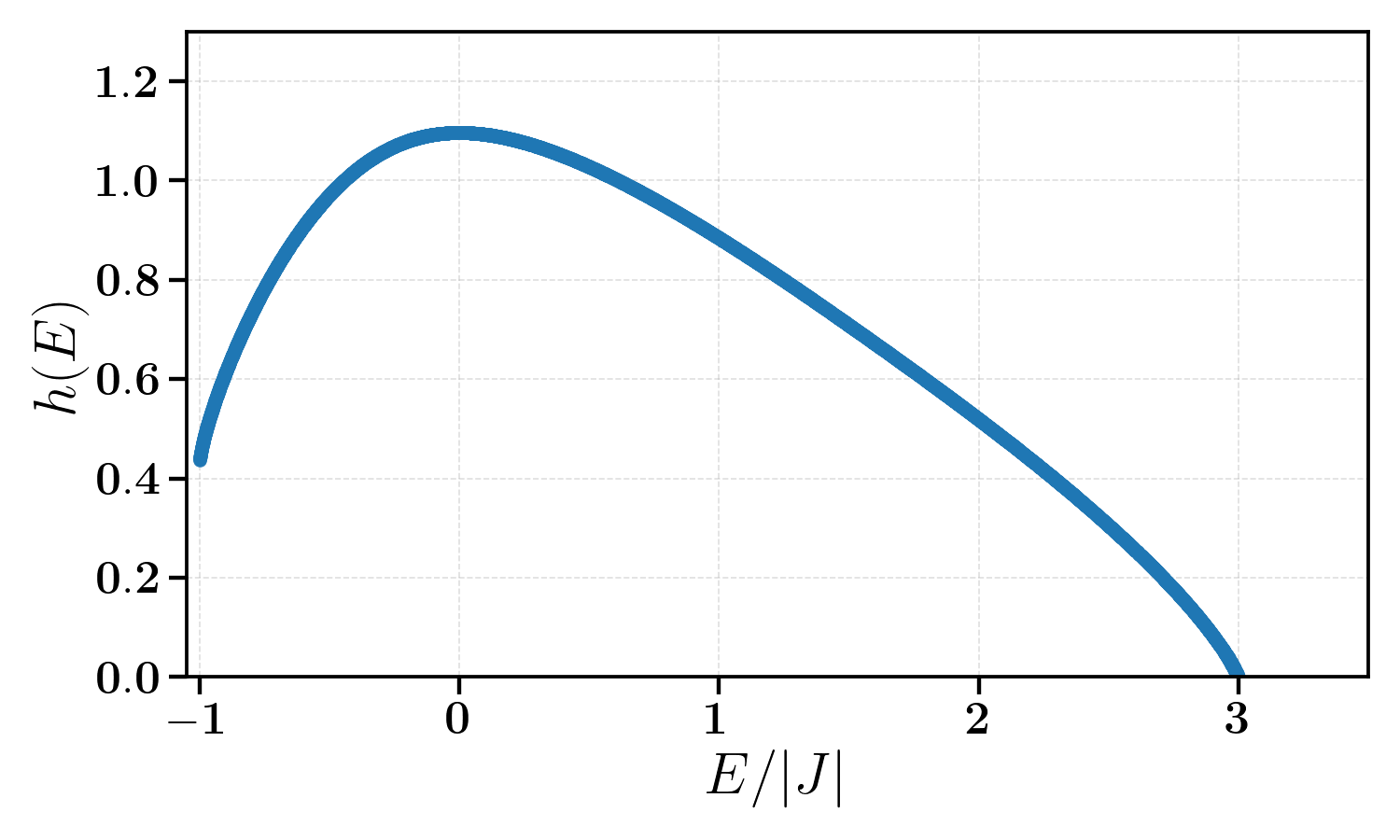}
	\caption{\label{fig:Tri_DOS} 
    The microcanonical entropy $h(E)=\ln g(E)/N$ per spin, where 
    $g(E)$ is the degeneracy of the system's level with the energy $E$ (per spin),
    in the spin-1 Ising model on the $47 \times 47$ triangular lattice.
    }
\end{figure}

We utilize the Wang-Landau (WL) algorithm \cite{WangLaudau:WL, WangLaudau:WL1}, augmented by the adaptive $1/t$ modification \cite{Belardinelli:t_Algor, Belardinelli:t_Algor1} to accurately determine the microcanonical entropy $h(E)=\ln g(E)/N$ per spin, where $E$ is the
degeneracy of the level with energy $E$, in the system (see Fig.~\ref{fig:Tri_DOS}).

The values of the ground-state entropy (per spin) in a system of the 
size $L\times L$ are shown in Fig.~\ref{fig:TriEntropy} for 
various $L=9\ldots 47$.
The illustrated data can be divided into three curves
corresponding to the three values of $L\, \text{mod}\, 3$,
i.e. to $L=3n$, $L=3n+1$ and $L=3n+2$, where $n$ is an integer.
Within each curve, the entropy is a monotonic function of $L$.
All three curves converge to the same value of entropy in the limit
of large $L$.

The obtained data indicate that the value $S_\infty$ of the entropy in the thermodynamic limit $L\rightarrow\infty$ is given by
\begin{equation}
\label{}
\begin{aligned}
S_{\infty} = 0.435854 \pm 0.000030 
\end{aligned}
\end{equation}
per spin.


\section{Dependence on the boundary conditions} 
\label{PBC explanation}

To accurately determine the ground-state entropy of the antiferromagnetic Ising model on the triangular lattice, it is crucial to adopt periodic boundary conditions (PBC) rather than free boundary conditions (FBC). This choice ensures translational symmetry across the lattice, thereby eliminating artificial edge effects that significantly alter the degeneracy structure of the system and compromise the accuracy of entropy calculations.

Specifically, the derivation of the partition function and the resulting analytical expression for the zero-temperature entropy~\cite{SedikSiyuSergey:Vacancy},
\begin{align}
    S_0 \left(0\right)=
    \nonumber\\
    \frac{1}{8\pi^2} \int_0^{2\pi} \int_0^{2\pi} \ln \left( 1 - 4 \cos \omega \cos\omega' + 4\cos^2\omega' \right) \, d\omega \, d\omega'
    \nonumber\\
        =  0.323066\ldots
    \label{ZeroPointEntropyClean}
\end{align}
explicitly relies on Fourier transformation techniques that inherently assume periodic boundary conditions. Without PBC, the lattice loses its translational invariance, invalidating momentum-space integration and precluding the exact recovery of Wannier's classical entropy result~\cite{Wannier:Ising,Wannier:erratum}.

Moreover, previous comparative studies have quantitatively demonstrated that free boundary conditions lead to lower entropy values due to edge-induced reductions in state degeneracy~\cite{Millane:Boundary}. Similarly, Nienhuis et al.~\cite{Nienhuis:TriangularSoS} emphasized that periodic boundary conditions are essential in accurately capturing the anisotropic properties of crystal shapes and properly describing critical phenomena, such as surface roughening transitions. Furthermore, Shevchenko et al.~\cite{Shevchenko:WangLandau} employed periodic boundary conditions in their Wang-Landau Monte Carlo simulations of diluted antiferromagnetic Ising models on frustrated lattices. Their approach enabled precise calculations of the residual entropy and revealed subtle dilution effects, underscoring the necessity of PBC in obtaining reliable thermodynamic properties in such systems. Therefore, we employ periodic boundary conditions throughout our simulations to ensure accurate, physically meaningful results consistent with established analytical predictions.

%

\end{bibunit}

\end{document}